\documentclass[preprint,superscriptaddress,showkeys,amsmath,amssymb,pra]{revtex4}

\usepackage{graphicx,amssymb}
\usepackage{dcolumn}
\usepackage{bm}

\newcommand{\beq}{\begin{equation}}
\newcommand{\eeq}{\end{equation}}
\newcommand{\bea}{\begin{eqnarray}}
\newcommand{\eea}{\end{eqnarray}}

\begin{document}

\title{Density functional theory of the trapped Fermi gas in the unitary regime}
\author{Brandon ~P. van Zyl}
\affiliation{Department of Physics, St. Francis Xavier University, Antigonish, Nova Scotia}
\author{D.~A.~W. Hutchinson}
\affiliation{The Jack Dodd Centre for Photonics and Ultra-Cold Atoms, Department of Physics, University of Otago, Dunedin, New Zealand}
\affiliation{Laboratoire Kastler Brossel, \'Ecole Normale Sup\'erieure, 24 rue Lhomond, 75231 Paris Cedex 05, France}
\author{Melodie Need}
\affiliation{Department of Physics, St. Francis Xavier University, Antigonish, Nova Scotia}


\begin{abstract}
We investigate a density-functional theory (DFT) approach for an unpolarized
trapped dilute Fermi gas in the unitary limit .  A reformulation of the
recent work of T. Papenbrock [Phys.~Rev.~A, {\bf 72}, 041602(R) (2005)] 
in the language
of fractional exclusion statistics allows us to
obtain an estimate of the universal factor, $\xi_{3D}$, 
in three dimensions (3D), in addition to providing 
a systematic treatment of finite-$N$ corrections.  We show that
in 3D,  finite-$N$ corrections lead to unphysical values for $\xi_{3D}$, 
thereby suggesting that a simple DFT applied to a small number of particles
may not be suitable in 3D.
We then perform an analogous calculation for the two-dimensional (2D) system
in the infinite-scattering length regime, and obtain a value of $\xi_{2D}=1$. 
Owing to the unique properties of the Thomas-Fermi energy 
density-functional in 2D our result, in contrast to 3D, is {\em exact} and 
therefore requires no finite-$N$ corrections.
\end{abstract}


\keywords{fermions, unitarity, two-dimensions}
\maketitle

\section{Introduction}

Harmonically trapped ultracold atomic Fermi gases have received considerable
attention within
the last few years owing to the experimental realization of Fermi degeneracy
by DeMarco and Jin~\cite{demarco}, the pioneering work of the Thomas group in the unitary regime~\cite{Ohara,Thomas} and other experimental efforts including, for example, Refs.~\onlinecite{chinc,chenq,regal,regal2}.  Using the phenomenon of a Feshbach
resonance, the two-body inter-particle interactions between the atoms
can be continuously 
tuned 
according 
to the magnitude of the magnetic field across the resonance region.  
Of particular interest is the so-called unitary regime which occurs at the
midpoint of this crossover,
and is characterized by the divergence of the scattering length due to the
existence of a zero-energy bound state for the two-body system~\cite{unitarity}.  
In this regime, the only relevant length scale is set by the
Fermi momentum, $k_f^{-1}$, and so the corresponding energy scale is the
Fermi kinetic energy.  Therefore, in three-dimensions (3D), 
the energy density should be proportional to that of a free Fermi gas
\beq
\frac{E}{N}=\xi_{3D} \left(\frac{E}{N}\right)_{\rm free} =
\xi_{3D} \frac{3k_f^2}{10 m}
\eeq
where $\xi_{3D}$ is a dimensionless proportionality constant, and
$k_f = (3\pi^2\rho)^{1/3}$.  In the unitary limit, the system
details do not contribute to the bulk thermodynamic 
properties~\cite{ho,heiselberg,baker}, and so
$\xi_{3D}$ is usually called the {\em universal factor}.  
Owing to the lack of a small expansion parameter, i.e., ($ak_f$), the 
exact determination of the universal factor $\xi_{3D}$ is a challenging
problem because the usual Green's function techniques for the many body system
are completely unreliable.  
Moreover, as the unitary Fermi gas may be relevant for the physics of 
neutron stars and high-$T_c$ superconductivity, the extraction of the
universal factor is also an important task.

To date, theoretical interest
in the universal factor  has 
been primarlily limited to
the 3D unitary Fermi gas
~\cite{ho,heiselberg,baker,engel,Papenbrock2005a,chen,svistunov,bulgac,carlson,note1}. 
These theoretical approaches vary in their level of sophistication, 
ranging from simple applications
of density-functional theory (DFT), to relativstic 
QED-like theories 
with hidden local symmetry, and quantum Monte Carlo (QMC) simulations. 
Unfortunately, the wide 
variety of theoretical approaches employed for the determination of 
$\xi_{3D}$ have also led to a broad range of possible values
for $\xi_{3D}$; namely, theory predicts $\xi_{3D} \sim 0.3 - 0.6$, although
Engelbrecht {\it et al}. have established an upper bound of $\xi_{3D} < 0.59$
from a BCS treatment~\cite{engel}.  
The disagreement between theoretical approaches not withstanding,
it is generally accepted that the recent 
QMC simulations
of Carlson {\em et al.}~\cite{carlson} yield the most reliable estimate
for the universal factor, which is $\xi_{3D} = 0.44 \pm 0.01$.

Initially it was difficult experimentally to obtain a value for $\xi_{3D}$, because
the precise location of the required Feshbach resonance was not known and a range of
estimates
$\xi_{3D} \sim 0.3 - 0.8$~\cite{zwierlein} were thus obtained.  With the position of the resonance now accurately determined, more recent experimental measurements of
$\xi_{3D}$ are in better agreement
and have yielded  
$\xi_{3D} = 0.51 \pm 0.04$~\cite{kinast} and $\xi_{3D} = 0.46 \pm 0.05$~\cite{part}.
These results strongly suggest that the QMC estimate of 
$\xi_{3D} \approx 0.44$ is indeed very reliable.

In this paper we determine 
the universal factor, $\xi$ for a trapped two-dimensional (2D) and 3D Fermi
gas.  
In 3D, our formulation of the problem in the language of fractional
exclusion statistics (FES) allows us not only to extract a value for
$\xi_{3D}$, but also provides a systematic treatment of finite-$N$
corrections.
Our determination of $\xi_{2D}$ is likewise obtained within an
exceedingly simple mathematical framework, which nevertheless leads to
an exact result for the universal factor in two-dimensions.
We are aware of no experimental results for the (quasi)
2D Fermi gas in, or near, the appropriate scattering regime and only two recent theoretical studies with the explicit aim of
extracting $\xi_{2D}$, both agreeing on
$\xi_{2D} = 1$~\cite{nussinov,nishida}. It has also been pointed out in
passing by Tonini {\em et al.}~\cite{yvan}
that for large scattering length in 2D, the effective interactions are weakly
attractive, thereby yielding the 
non-interacting Fermi gas in the infinite scattering length limit~\cite{unitarity}; this observation plays a critical role in our subsequent determination
of $\xi_{2D}$.

\section{Trapped Fermi gas in the unitary regime}

In this section, we apply a simple DFT in order to determine the
universal factor for an unpolarized trapped Fermi gas. 
We begin by reformulating the work of Papenbrock in 3D, with
particular attention paid to
finite-$N$ corrections, and what impact they have on the
estimate of $\xi_{3D}$.   We also present an analogous
analysis in 2D, and show that it  yields an exact result for $\xi_{2D}$
without requiring any finite-$N$ corrections.

\subsection{Three Dimensions}

As stated in the introduction, Carlson {\it et al.} have performed 
simulations for a uniform 3D Fermi gas in the unitary regime.  Dimensional
arguments suggest that in this regime, the energy must be proportional to that
of a free Fermi gas, 
\beq
E(N) = \xi_{3D} E_{\rm free}(N)~,
\label{eunitary}
\eeq
where $E_{\rm free}(N)$ is the Thomas-Fermi (TF) energy of $N$ noninteracting 
spin-$1/2$ fermions.  Surprisingly, the QMC simulations with
$N =10 - 40$ fermions agree very well with Eq.~(\ref{eunitary}) in spite
of the fact that the exact quantum mechanical energies are known to
differ from the corresponding TF energies due to shell and finite size
effects.  Motivated by this result, and again
on dimensional grounds,
Papenbrock~\cite{Papenbrock2005a} has suggested that a density functional from
TF theory, 
\beq
\varepsilon^{\rm 3DHO}_{\rm TF}(\rho)=\xi_{3D} \frac{\hbar^2}{m}c\rho^{5/3}
+ \frac {1}{2}m \omega ^2 r^2 \rho~,
\label{3dDFT}
\eeq
should be a good approximation to the exact density functional.  In general,
gradient corrections should be included, but Eq.~(3) represents the simplest 
possible form for the energy density-functional.
In the above, $\omega$ is the trapping frequency of an 
isotropic harmonic oscillator (HO) potential, 
and $c=\frac{3}{10}(3\pi^2)^{2/3}$.
It is straightforward to show that a minimization of this functional, 
subject to a fixed number of particles, leads to
\beq
\rho ^{\rm 3DHO}_{\rm TF} = \frac {1}{3 \pi ^2} \left( \frac {2}{\xi_{3D}
l^2}\right)^{3/2} \left( (3N)^{1/3} \xi_{3D}^{1/2} - \frac{r^2}{2l^2}\right)^{3/2}~,
\eeq
where $l=(\hbar/m \omega)^{1/2}$ is the oscillator length. Integrating this
energy functional up to the TF radius yields the total TF energy
\beq
E_{\rm TF}^{\rm 3DHO}= \frac{1}{4}(3N)^{4/3}\xi_{3D}^{1/2}\hbar \omega.
\eeq
The energy can be calculated exactly for the case $N=2$, 
where there is a binding
energy of $\hbar \omega$ in the unitary limit such that total energy for two
fermions in a harmonic trap
$E_{\rm ex} = 2 \hbar \omega$~\cite{Busch1998a}. Comparison with the
TF expression (which is expected to be reasonable for all even numbers of
fermions) with $N=2$ gives a value for the universal constant of 
$\xi_{3D} \approx
0.54$.  Although it is not explicitly 
mentioned in Ref.~\onlinecite{Papenbrock2005a}, if
one uses the exact two-particle density
\beq
\rho_{\rm ex}(r) = \frac{4}{\pi^{3/2}l^3}\frac{l}{r}e^{-2(r/l)^2}
\int_{0}^{r}dx e^{x^2}~,
\eeq
in Eq.~(3) rather than Eq.~(4), and then performs the spatial integration, 
a much better value of $\xi_{3D} \approx 0.48$ is obtained.  Thus, 
a very simple DFT already yields a value for $\xi_{3D}$ that is within 8 \%
of the generally accepted QMC result.  
It is then natural to ask what effect finite-$N$ corrections will have on
this simple result; afterall, TF DFT is generally only valid for $N \gg 1$,
and/or slowly spatially varying potentials, which
is certainly not the case in the present analysis.
In order to investigate the finite-$N$ corrections, it turns out to
be more convenient to reformulate
the DFT of Ref.~\onlinecite{Papenbrock2005a} in a slightly different
manner.  

Bhaduri, {\em et al.}~\cite{bhaduri} have argued that the uniform 3D interacting Fermi gas at unitarity can be described in terms of a non-interacting gas obeying fractional exclusion statistics (FES).
This generalization of the standard Bose-Einstein and Fermi-Dirac particle statistics 
was developed by Haldane~\cite{haldane} in the context of the quasiparticle excitations 
of the fractional quantum Hall effect, and later generalized by Wu~\cite{wu}. In these Haldane-Wu statistics the concept of the Pauli exclusion principle is extended such that the distribution function for a single particle state with energy $\varepsilon$ is given by
\beq
n(\varepsilon)=\frac{1}{w+\alpha}~,
\eeq
where the temperature dependent function $w$ is defined through
\beq
w(1+w)^{1-\alpha}=e^{\beta(\varepsilon - \mu)}.
\eeq
In the limit $\alpha=0$ this then reduces to the usual Bose distribution and for $\alpha=1$, the Fermi-Dirac distribution. At zero temperature the distribution reduces to
\bea
n(\varepsilon)= \frac{1}{\alpha}, \hspace{3mm} \varepsilon &<& \mu \nonumber\\
n(\varepsilon)= 0, \hspace{3mm} \varepsilon &>& \mu.
\eea
The parameter $\alpha$ that defines the statistics gives a modified exclusion principle which states that the maximum occupancy of a single state is $1/\alpha$ -- hence fractional exclusion statistics.

Let us now consider
the density of states, $D(\varepsilon)$, for the 3D HO (including the spin
degeneracy factor of 2)~\cite{semiclassical} 
\beq
D(\varepsilon) = \frac{\varepsilon^2}{(\hbar\omega)^3} -
\frac{1}{4(\hbar\omega)}~.
\eeq
Following Bhaduri {\it et al.}~\cite{bhaduri}, we now utilize the Haldane-Wu statistics 
to write the total number of particles, $N$, in the system as
\beq
N = \frac{1}{\alpha}\int_{0}^{\mu}D(\varepsilon)~d\varepsilon~,
\eeq
where, as above, $1/\alpha$ replaces the usual
zero temperature occupancy factor of unity for fermions.
Rescaling all energies with respect
to the oscillator energy $\hbar\omega$ gives the following expression
for the chemical potential upon integration
\beq
\frac{\mu^3}{3} - \frac{\mu}{4} - (\alpha N) = 0~.
\eeq
This equation has one physically acceptable root, viz.,
\begin{eqnarray}
\mu &=& \frac{1}{2}\left(12\alpha N + \sqrt{144\alpha^2N^2-1}\right)^{1/3} +
\frac{1}{2}\frac{1}{(12\alpha N+\sqrt{144\alpha^2N^2-1})^{1/3}}\nonumber \\
&=& (3\alpha N)^{1/3} + \frac{1}{4}\frac{1}{(3\alpha N)^{1/3}} - 
\frac{1}{192}\frac{1}{(3\alpha N)^{5/3}} +\cdot\cdot\cdot
\end{eqnarray}
to $O(N^{-5/3})$ corrections.
Similarly, the total energy of the system is given by
\begin{eqnarray}
E &=& \frac{1}{\alpha }\int_0^{\mu}D(\varepsilon)\varepsilon~d\varepsilon\nonumber \\
&=& \frac{1}{\alpha}\left(\frac{\mu^4}{4} - \frac{\mu^2}{8}\right)~. \label{finiteN}
\end{eqnarray}
Rewriting Eq.~(\ref{finiteN}) explicitly in terms of $N$ results in the expression
(to $O(N^{-2/3})$ corrections) 
\beq
E = \alpha^{1/3}\frac{(3N)^{4/3}}{4} + \alpha^{-1/3}\frac{(3N)^{2/3}}{8} +\alpha^{-1} \frac{1}{32} +
\alpha^{-5/3}\frac{1}{384}\frac{1}{(3N)^{2/3}} + \cdot\cdot\cdot
\eeq
We can now make a formal connection with Ref.~\onlinecite{Papenbrock2005a} 
by identifying
$\xi_{3D} = \alpha^{2/3}$.  Retaining
only the first term in Eq.~(15), we then obtain 
\beq
E = \frac{1}{4}(3N)^{4/3}(\xi_{3D})^{1/2}~,
\eeq
which is identical to the result for the total TF
energy given in \cite{Papenbrock2005a}.
For the $N=2$ case, we have $E_{\rm ex} = 2$, and we obtain as a first
approximation to $\alpha$
\beq
2 = \alpha^{1/3}\frac{6^{4/3}}{4} \approx 2.725 \alpha^{1/3}~,
\eeq
which immediately gives $\alpha=32/81$, or 
$\xi_{3D} \approx 0.54$, in complete agreement with
the DFT approach of Papenbrock. 
If we keep the first leading order correction in Eq.~(15), we obtain
(again, for the $N=2$ case) a second approximation to $\alpha$
\beq
2 = \alpha^{1/3}\frac{6^{4/3}}{4} + \alpha^{-1/3}\frac{6^{2/3}}{8}~,
\eeq
which has no real roots.  
However, if we retain the first and third (i.e., $\alpha^{-1}/32$)
terms in Eq.~(15), we obtain 
a much improved value of $\xi_{3D} \approx 0.49$.  
Nevertheless, beyond the first and this latter approximation,
we find that there is no real value of $\alpha$ which yields 
$E=2$.
Thus, a systematic treatment of the finite-$N$
corrections (sometimes referred to as the extended Thomas-Fermi theory (ETF)),
with the goal of yielding an improved value for
$\xi_{3D}$, fails in 3D (for the $N=2$ case at least). Of
course this could also be an indication that the identification of the
strongly interacting Fermi gas with a gas of non-interacting particles
obeying FES is not valid (in 3D) for small particle numbers once finite-{\em N} corrections are
considered. Bhaduri {\it et al.}~\cite{bhaduri} have however used this approach, even for small numbers of particles, to obtain thermodynamic properties of the unitary gas at finite-temperature which compare well with the Monte Carlo results~\cite{svistunov,bulgac}. It is also worth noting that for $\alpha=1$ (i.e., Fermi-Dirac statistics),
the exact energy and chemical potential are $E=3$ and $\mu=2$, respectively.
In this case, the systematic corrections discussed above do yield improved
values.  For example, the first approximation gives $E=2.73$ and $\mu=1.82$,
while the second approximation already gives $E=3.14$ and $\mu=1.96$.
This said, the fact that the first approximation, and its augmented form (i.e.,
including the $N$-independent term) yield values
for the universal factor so close to the QMC result is perhaps fortuitous.
Interestingly, the requirement that $E$ be real leads to a
{\em lower bound} of $\xi_{3D} > 0.12$, and 
the energy takes on a minimum value $E\approx 2.45$ for $\xi_{3D} \approx 0.3$.

\subsection{Two Dimensions}

Turning now to the two-dimensional isotropic harmonic trap, the TF density functional reads (for the noninteracting Fermi gas)
\beq
\varepsilon_{\rm TF}^{\rm 2DHO}(\rho)= \frac {\hbar^2 \pi}{2 m} \rho^2 +
\frac{1}{2}m \omega^2 r^2 \rho. \label{exact}
\eeq
At zero temperature, when filling $M+1$ oscillator shells, 
Brack and van Zyl~\cite{brandon} have provided
an {\em exact}, closed form solution for the 
zero-temperature particle density in 2D, 
\beq
\rho(r) = \frac{2m\omega}{\pi\hbar} \sum_{n=0}^{M}(M-n+1) (-1)^n L_n
\left(\frac{2m\omega}{\hbar}r^2\right)e^{\frac{m\omega}{\hbar}r^2}~, \label{rho}
\eeq
where $L_n(x)$ is a Laguerre polynomial.
Inserting Eq.~(\ref{rho}) into Eq.~(\ref{exact}), and integrating
over all space yields the total TF energy
\beq
E^{\rm 2DHO}_{\rm TF} = \frac{\hbar\omega}{3}[2M^3 + 9M^2 + 13M + 6]~. \label{QMT}
\eeq
The number of particles in $M+1$ filled shells is given by
$N(M) = M^2 + 3M +2$.  What is truly remarkable about Eq.~(\ref{QMT}) is that it
is also the {\em exact} quantum mechanical energy of the system.  In other words,
the TF functional (which is exact only in the homogeneous
limit), {\em without gradient corrections} yields
the exact total energy of the inhomogeneous noninteracting system.  

Before applying the above result to the trapped 2D Fermi gas at unitarity,
we must first make clear what is meant by ``unitarity'' in two-dimensions.
If one considers the 2D case (see our comments for 3D in \cite{unitarity})
the $s$-wave scattering cross-section takes the form
($\hbar=2m=1$)
\beq
\sigma_{2D} = \frac{4\pi^2}{k\left\{\pi^2+4\left[{\rm ln}
\left(\frac{1}{a_{2D}k}\right)\right]^2\right\}}~,
\eeq
with $a_{2D}$ now the 2D scattering length.  The maximal value for the
cross-section is then obtained when the logarithmic term vanishes.  
Following the equivalent argument to that in 3D, the unitarity limit
is therefore obtained when $\frac{1}{a_{2D}k}\rightarrow 1$.  In
general, this is not what is taken as the unitary limit in the
literature.  For example, Refs.~\onlinecite{nussinov} and
\onlinecite{nishida} refer to the limit $a_{2D} \rightarrow \infty$.
Clearly, if we hold that unitarity occurs when $\sigma_{2D}$ reaches
is maximal value, the dimensional arguments leading to fact that the energy of
the interacting gas must be proportional to the energy of the noninteracting
gas are not applicable.  In particular, the hierarchy:
$a \gg k_f^{-1}\gg r_0$, implicity used in obtaining Eq.~(2) for 3D, 
is not sufficient in 2D as the range of the potential, $r_0$,
{\em cannot generally be taken to be zero}.  Indeed, unitarity in 2D implies
that $a_{2D} = k_f^{-1}$.  Thus, if we wish to preserve an analogous notion of
unitarity in 2D, we require that the 2D scattering length become
very large such that $a_{2D}k_f \gg 1$.  Obviously, this does not
correspond to a maximum in $\sigma_{2D}$, but the large
(i.e., essentially infinite) scattering length limit is
what is often regarded as the unitary regime in two-dimensions
~\cite{nussinov,nishida}.

With the above qualification regarding unitarity in 2D in mind, we now consider
the binding energy of a dimer in free space, $E_0$~\cite{yvan},
\beq
E_0 = \frac{4\hbar^2}{ma_{2D}^2e^{2\gamma}}
\eeq
where $\gamma = 0.57721.....$ is Euler's constant, along with the
coupling constant for $s$-wave scattering between two particles, 
\beq
\frac{1}{g_0} = \frac{m}{2\pi\hbar^2}
\left[
{\rm log}\left(\frac{r_0}{\pi a_{2D}}\right) - \gamma + \frac{2G}{\pi}\right]~,
\eeq
where $G=0.91596....$ is Catalan's constant.  In the limit
$a_{2D} \rightarrow \infty$, for a fixed density and range $r_0$,
$g_0\rightarrow 0^{-}$ and $E_0 \rightarrow 0^{+}$.  Thus, in the
limit $a_{2D}k_f \gg 1$, the trapped 2D gas corresponds to a weakly
attractive Fermi gas with a zero energy bound state.  In other words,
the threshold for the appearance of the first zero-energy bound state
corresponds to zero coupling.  In this sense, the infinite scattering length limit
in 2D corresponds to a noninteracting Fermi gas, which is in stark
contrast to 3D, where the $(a_{3D}k_f)\gg 1 $ regime requires a non-trivial
treatment of interactions.

We are now in a position to apply the simple DFT approach to the
trapped 2D Fermi gas in this limit.
Arguing as in Ref.~\onlinecite{Papenbrock2005a}, the appropriate 2D TF 
density functional is given by
\beq
\varepsilon_{\rm TF}^{\rm 2DHO}(\rho)=\xi_{2D} \frac {\hbar^2 \pi}{2 m} \rho^2 +
\frac{1}{2}m \omega^2 r^2 \rho.
\eeq
Again, gradient corrections generally have to be introduced, 
but as we will show, they
are not required for the 2D case.
Minimizing with respect to $\rho$, subject to the constraint that the 
total number of particles is held fixed, yields a chemical potential
\beq
\mu_{\rm TF} ^{\rm 2DHO}=\xi_{2D}^{1/2} N^{1/2} \hbar \omega
\eeq
with TF density
\beq
\rho_{\rm TF} ^{\rm 2DHO}= \frac {m}{\xi_{2D} \hbar^2 \pi} \left(\mu_{\rm TF} ^{\rm 2DHO} - \frac{1}{2}m
\omega^2 r^2\right). \label{den_2d}
\eeq
Inserting Eq.~(\ref{den_2d}) into the energy functional and integrating 
to the TF radius
yields a total TF energy 
\beq
E_{\rm TF}^{\rm 2DHO}= \frac{2}{3}N^{3/2}\xi_{2D}^{1/2}\hbar \omega. \label{old_22}
\eeq
As we have already discussed, in
the infinite scattering length regime the exact energy for two fermions 
in a harmonic trap is zero energy
bound state. 
Hence the exact energy for a pair of fermions in a 2D 
harmonic trap is also $E_{\rm ex} = 2 \hbar \omega$. 
Comparison of the TF energy, viz., (\ref{old_22}) with Eq.~(21), which
is exact in the infinite scattering length limit,
gives a value 
for $\xi_{2D}$ such that~\cite{1D_note}
\beq
\xi_{2D} = 1 + \frac{1}{4N}
\label{wrong}
\eeq
For example, with $N=2$, we obtain $\xi_{2D} = 9/8=1.125$.  
Thus, while an analysis of finite-$N$ corrections in 2D
is not applicable here,
because $D(\varepsilon) = \varepsilon/(\hbar\omega)^2$,  
Eq.~(29) illustrates that employing the 2D TF density in Eq.~(25) 
leads to (what turns out to be) an {\em overestimate} of $\xi_{2D}$, with
$\xi_{2D}\rightarrow 1^{+}$ as $N\rightarrow\infty$.
%
%

We have, however already established that Eq.~(\ref{exact}) yields the exact
quantum mechanical energy provided the exact density is used as input.  Since
the 2D gas in the infinite scattering length limit corresponds to the noninteracting system, if
we use the exact $N$-particle density, instead of the TF density, viz.,
Eq.~(\ref{den_2d}), and then perform the spatial integration of (\ref{exact}), we 
immediately obtain $\xi_{2D} = 1$ identically.  This is true for
any $N = M^2+3M+2$, and thus establishes that the simple 2D TF functional
does indeed lead to a universal $\xi_{2D}=1$, independent of the particle
number, when the exact density is used.  Recall that in the 3D case,
the use of the exact two-particle density, Eq.~(6), does
lead to an improved value of $\xi_{3D}\approx 0.48$ as compared to that obtained using the TF density.
However, as the 3D TF functional is 
{\em not} exact in the unitary regime, neither is the extracted universal 
factor.
%

\section{Connection with FES in the 2D Fermi gas.}
In a relatively recent paper, Srivastava {\it et al.} have considered a
gas of $N$ fermions interacting via a repulsive 
two-body zero-range (psuedo) potential
within a mean-field approach~\cite{note3}.  Specifically, they showed that 
within the TF formalism,
the zero temperature density functional (with a spin-degeneracy factor of
2 included) is given by
\beq
\varepsilon_{\rm FES}^{\rm 2DHO} = \alpha \frac{\hbar^2}{2m}\pi\rho^2
+ \frac{1}{2}m\omega^2r^2\rho~,
\eeq
where $\alpha = (1+\frac{mM_0}{2\pi\hbar^2})$ is a dimensionless statistical
parameter, 
and $M_0$ is the zeroth moment of the two-body potential.
The above equation implies that a gas of fermions in 2D, interacting by
a repulsive two-body zero-range potential, appear as particles obeying
ideal FES (at least within the TF framework).  Notice that the form of Eq.~(30) is
identical to Eq.~(25), but here, $\alpha$ is a statistical parameter
unreliant upon being in the unitary or
infinite scattering length regimes.  In fact, the scaling of the
kinetic energy functional here relies entirely on the details of the two-body
potential, whereas in the unitary limit, the details of the potential
are unimportant.
The similarities between Eqs.~(25) and (30), and the analgous expressions
for $\rho$, $\mu$ and $E$, naturally lead to the question
of what connection the Fermi gas in the infinite scattering length limit has with FES.  Unfortunatley, the
answer is quite anticlimactic in 2D because it is only in the
limit $\alpha =1$, that there exists any
correspondence.
This, unsurprisingly,  
leads immediately to $\xi_{2D} = 1$ since 
$\alpha = \xi_{2D} + \frac{mM_0}{2\pi\hbar^2}$.

\section{Conclusions}
In this paper we have reformulated the simple DFT of Papenbrock in the
language of fractional exclusion statistics to obtain an 
estimate of the universal factor $\xi_{3D}$ which is in good agreement with 
that obtained via QMC 
simulations, and experiments.   We have shown, however, that trying to go beyond this estimate by
using a systematic treatment of finite $N$ corrections fails in 3D.

In contrast, when we apply the DFT technique to the 2D trapped Fermi gas in the infinite scattering length limit (and use the exact density in the 2D TF energy density functional), we have shown that the {\em exact} quantum mechanical energy is obtained for 
any number of closed shells. The TF result thus obtained is therefore not limited to $N \gg 1$ as in the 3D case and we are able to establish that the universal factor for the harmonically trapped 2D Fermi gas is exactly $\xi_{2D}=1$.   
We therefore conclude that
a simple DFT approach, as originally proposed by Papenbrock, is only strictly valid 
in 2D, where the TF energy functional, Eq.~(19), 
has the remarkable property of being
exact, {\em without gradient corrections}.


\begin{acknowledgments}
We would like to thank the NSERC of Canada, the Marsden Fund of New Zealand, and C.N.R.S. of France for fiscal support. For their warm hospitality, DAWH would like to thank the cold atoms group at ENS where this work was concluded and Yvan Castin in particular for useful discussions.
MN would like to acknowlege an NSERC USRA award for additional financial
support.  We are also indebted to R. K. Bhaduri for suggesting the
investigation of finite-$N$ corrections presented in Sec.~II(A)
of this paper.
\end{acknowledgments}


\begin{thebibliography}{90}
\bibitem{unitarity}
In 3D the unitary limit is defined as the point of maximal scattering cross-section. Setting here $\hbar=2m=1$ the 3D scattering cross-section takes the 
form (for $s$-wave scattering)
\beq
\sigma_{3D} = \frac{4 \pi}{\left(1/a_{3D}\right)^2 + k^2}, \nonumber
\eeq
where $a_{3D}$ is the 3D scattering length. This clearly takes a maximal value for all wavevectors $k$ as $a_{3D} \rightarrow \pm \infty$.
\bibitem{note1} In actual experiments, the gas is also trapped by an optical
potential $V(r)$, and is thus inhomogeneous.  Contact with experiments
in this situation is usually facilitated by the use of the local-density
approximation.
\bibitem{1D_note} A similar calculation can be performed for the one dimensional trapped Fermi gas in the infinite scattering length limit. In this case the Thomas-Fermi energy is given by
\beq
E^{\rm 1DHO}_{\rm TF}= \frac{1}{4} N^2 \xi_{1D}^{1/2} \hbar \omega \nonumber
\eeq
and comparing to the exact result for $N=2$ gives a value for the analogous ``unitary'' factor, $\xi_{1D}$, of unity.  This happens to be also the
{\em exact} result in 1D, but this is fortuitous because it is known
that the TF functional in 1D  is not exact, similar to the situation found
in 3D.  See, e.g., Ref.~\onlinecite{brandon}
\end{thebibliography}
\end{document}